\documentclass{ws-procs975x65}

\begin{document}

\title{CONSTRUCTIVE PROOF OF THE KERR-NEWMAN\\ BLACK HOLE UNIQUENESS:\\
DERIVATION OF THE FULL SOLUTION FROM SCRATCH}

\author{REINHARD MEINEL and REN{\'E} RICHTER}

\address{University of Jena, Theoretisch-Physikalisches Institut,\\
Max-Wien-Platz 1, 07743 Jena, Germany\\
\email{meinel@tpi.uni-jena.de}}

\begin{abstract}
The Kerr-Newman black hole solution can be constructed
straightforwardly as the unique solution to the boundary value problem of the
Einstein-Maxwell equations corresponding to an asymptotically flat, stationary
and axisymmetric electro-vacuum spacetime surrounding a connected Killing
horizon.
\end{abstract}

\bodymatter

\section{Introduction}
It was shown in Ref.~\refcite{m12} that the complex Ernst potentials\cite{ernst68} 
on the symmetry axis of an asymptotically flat, stationary and axisymmetric 
electro-vacuum spacetime surrounding a connected Killing horizon (including the 
case of a degenerate horizon) can be constructed uniquely by means of the inverse 
scattering method\cite{bv01, rfe}. This provides a new proof of the Kerr-Newman 
black hole uniqueness since the full solution can straightforwardly be derived 
from the axis data, cf.~Ref.~\refcite{he81}. Here we present a particular method 
for obtaining the Ernst potentials off the axis, see Sec.~4. 

\section{Ernst equations and Linear Problem}
The stationary and axisymmetric Einstein-Maxwell vacuum equations are equivalent 
to the Ernst equations\cite{ernst68}
\begin{equation}
f\,\Delta\, {\mathcal E}=(\nabla {\mathcal E} +
2\bar{\Phi}\nabla \Phi)\cdot\nabla {\mathcal E}\, , \quad
f\,\Delta\, \Phi=(\nabla {\mathcal E} +2\bar{\Phi}\nabla \Phi)\cdot\nabla \Phi
\end{equation}
\begin{equation}
\mbox{with} \quad f\equiv\Re\, {\mathcal E} + |\Phi|^2\, , \quad 
\Delta=\frac{\partial^2}{\partial\rho^2}+\frac{1}{\rho}\frac{\partial}{\partial\rho}
+\frac{\partial^2}{\partial\zeta^2} \, , \quad \nabla
=(\frac{\partial}{\partial\rho},\frac{\partial}{\partial\zeta})\, ,
\end{equation}
where $\rho$ and $\zeta$ are Weyl coordinates and a bar denotes complex conjugation. 
These equations are the integrability condition of a related Linear Problem 
(LP)\cite{a80, nk83}:
\begin{equation}
{\bf Y}_{,z}=\left[\left(\begin{array}{ccc}
B_1 & 0 & C_1 \\
0 & A_1 & 0 \\
D_1 & 0 & 0
\end{array}
\right)+\lambda\left(\begin{array}{ccc}
0 & B_1 & 0 \\
A_1 & 0 & -C_1 \\
0 & D_1 & 0
\end{array}
\right)\right]{\bf Y}\, ,
\end{equation}
\begin{equation}
{\bf Y}_{,\bar{z}}=\left[\left(\begin{array}{ccc}
B_2 & 0 & C_2 \\
0 & A_2 & 0 \\
D_2 & 0 & 0
\end{array}
\right)+\frac{1}{\lambda}\left(\begin{array}{ccc}
0 & B_2 & 0 \\
A_2 & 0 & -C_2 \\
0 & D_2 & 0
\end{array}
\right)\right]{\bf Y}
\end{equation}
\begin{equation}
\mbox{with} \quad \lambda=\sqrt{\frac{K-{\rm i}\bar z}{K+{\rm i}z}}\, ,
\quad
z=\rho+{\rm i}\zeta\, , \quad \bar z=\rho-{\rm i}\zeta\, ,
\end{equation}
\begin{equation}
A_1=\bar{B}_2=\frac{{\mathcal E}_{,z}+2\bar{\Phi}\Phi_{,z}}{2f}\, , \quad
A_2=\bar{B}_1=\frac{{\mathcal E}_{,\bar z}+2\bar{\Phi}\Phi_{,\bar z}}{2f}\, ,
\end{equation}
\begin{equation}
C_1=f\bar{D}_2=\Phi_{,z}\, , \quad C_2=f\bar{D}_1=\Phi_{,\bar z}\, ,
\end{equation}
where $K$ is the complex ``spectral parameter''. 

\section{Solution on the axis} 
On the upper part of the symmetry axis ($\rho=0$, $\zeta>l$), denoted by 
$\mathcal{A^+}$, a suitably normalized solution to the LP for the black hole 
problem (horizon at $\rho=0$, $|\zeta|\le l$) reads\cite{m12}
\begin{equation}\label{Y+}
\mathcal{A^{+}}: \quad {\bf Y}_{+}=
\left(\begin{array}{crr}
\bar{\mathcal E}_++2|\Phi_+|^2 & 1 & \Phi_+ \\
{\mathcal E_+} &-1 & -\Phi_+ \\
2\bar \Phi_+ & 0 & 1
\end{array}
\right)
\left(\begin{array}{ccc}
F & 0 & 0 \\
G & 1 & L \\
H & 0 & 1
\end{array}\right)\, ,
\end{equation}
where we have assumed $\lambda=+1$ for $K\neq\zeta$. The Ernst potentials 
$\mathcal E_+(\zeta)$ and $\Phi_+(\zeta)$ on $\mathcal{A^+}$ are given by
\begin{equation}
\mathcal E_+=1-\frac{2M}{\zeta+M-{\rm i}J/M} \, , \quad 
\Phi_+=\frac{Q}{\zeta+M-{\rm i}J/M} 
\end{equation}
together with the parameter relation
\begin{equation}\label{P1}
\frac{l^2}{M^2}+\frac{Q^2}{M^2}+\frac{J^2}{M^4}=1\, ,
\end{equation}
where $M$, $J$ and $Q$ denote mass, angular momentum and charge. The functions 
$F(K)$, $G(K)$, $H(K)$ and $L(K)$ are given by
\begin{equation}
F=\frac{(K-L_1)(K-L_2)}{(K-K_1)(K-K_2)}\, , \quad G=\frac{Q^2-2{\rm i}J}{(K-K_1)(K-K_2)}\, ,
\end{equation}
\begin{equation}
H=-\frac{2Q(K-L_1)}{(K-K_1)(K-K_2)}\, , \quad L=-\frac{Q}{K-L_1}
\end{equation}
\begin{equation}
\mbox{with} \quad L_{1/2}=-M\pm{\rm i}\frac{J}{M}\, , \quad
K_{1/2}=\pm\sqrt{M^2-Q^2-\frac{J^2}{M^2}}\, . 
\end{equation}

\section{Solution off the axis}
Using the relation\cite{m12}
\begin{equation}
{\bf Y}(\rho,\zeta,-\lambda)=\left(\begin{array}{crc}
1 & 0 & 0 \\
0 & -1 & 0 \\
0 & 0 & 1
\end{array}
\right){\bf Y}(\rho,\zeta,\lambda)
\left(\begin{array}{ccc}
0 & 1 & 0 \\
1 & 0 & 0 \\
0 & 0 & 1
\end{array}\right)
\end{equation}
together with
\begin{equation}\label{Y1}
{\bf Y}(\rho,\zeta,1)=
\left(\begin{array}{crr}
\bar{\mathcal E}+2|\Phi|^2 & 1 & \Phi \\
{\mathcal E} &-1 & -\Phi \\
2\bar \Phi & 0 & 1
\end{array}
\right)
\end{equation}
we are led from \eqref{Y+} to the following ansatz for ${\bf Y}(\rho,\zeta,\lambda)$:
\begin{equation}
{\bf Y}(\rho,\zeta,\lambda)=\left(\begin{array}{crc}
\psi(\rho,\zeta,\lambda) & \psi(\rho,\zeta,-\lambda) & \alpha(\rho,\zeta,\lambda) \\
\chi(\rho,\zeta,\lambda) & -\chi(\rho,\zeta,-\lambda) & \beta(\rho,\zeta,\lambda) \\
\varphi(\rho,\zeta,\lambda) & \varphi(\rho,\zeta,-\lambda) & \gamma(\rho,\zeta,\lambda)
\end{array}
\right)
\end{equation}
with
\begin{equation}
\psi=1+a_1\left(\frac{1}{\kappa_1-\lambda}-\frac{1}{\kappa_1+1}\right)
+a_2\left(\frac{1}{\kappa_2-\lambda}-\frac{1}{\kappa_2+1}\right)\, ,
\end{equation}
\begin{equation}
\chi=1+b_1\left(\frac{1}{\kappa_1-\lambda}-\frac{1}{\kappa_1+1}\right)
+b_2\left(\frac{1}{\kappa_2-\lambda}-\frac{1}{\kappa_2+1}\right)\, ,
\end{equation}
\begin{equation}
\varphi=c_1\left(\frac{1}{\kappa_1-\lambda}-\frac{1}{\kappa_1+1}\right)
+c_2\left(\frac{1}{\kappa_2-\lambda}-\frac{1}{\kappa_2+1}\right)\, ,
\end{equation}
\begin{equation}
\alpha=\Phi+\frac{\alpha_0}{K-L_1}\, , \quad 
\beta=-\Phi\,\frac{\lambda(K+{\rm i}z)}{K-L_1}\, ,
\quad \gamma=1+\frac{\gamma_0}{K-L_1}\, ,
\end{equation}
\begin{equation}
\mbox{where} \quad \kappa_i=\sqrt{\frac{K_i-{\rm i}\bar z}{K_i+{\rm i}z}}
\quad (\mathcal{A^+}:\quad \kappa_i=+1)\, .
\end{equation}
A careful discussion of the regularity conditions at 
\begin{equation}
\lambda_i=\sqrt{\frac{L_i-{\rm i}\bar z}{L_i+{\rm i}z}}
\end{equation}
leads to a set of linear algebraic equations which uniquely determine the 
unknowns $a_i(\rho,\zeta)$, $b_i(\rho,\zeta)$, $c_i(\rho,\zeta)$, 
$\alpha_0(\rho,\zeta)$, $\gamma_0(\rho,\zeta)$ and $\Phi(\rho,\zeta)$. 
According to \eqref{Y1}, ${\mathcal E}(\rho,\zeta)$ is given by 
$\chi(\rho,\zeta,1)$. The resulting Ernst potentials are the well-known 
expressions
\begin{equation}
\mathcal E=1-\frac{2M}{r-{\rm i}(J/M)\cos \theta} \, , \quad 
\Phi=\frac{Q}{r-{\rm i}(J/M)\cos \theta} 
\end{equation} 
with Boyer-Lindquist coordinates 
$r$ and $\theta$ related to our Weyl coordinates $\rho$ and $\zeta$ by
\begin{equation}
\rho=\sqrt{r^2-2Mr+J^2/M^2+Q^2}\,\sin\theta \, , \quad
\zeta=(r-M)\cos\theta \, .
\end{equation}

\end{document}